# Noise characteristics of 3D time-of-flight cameras

Dragos Falie, Vasile Buzuloiu
dfalie@alpha.imag.pub.ro
Polytechnic University of Bucuresti, the Image Processing and Analysis Laboratory, Romania

*Abstract*—**Time-of-flight (TOF) cameras are based on a new technology that delivers distance maps by the use of a modulated light source. In this paper we first describe a set of experiments that we performed with TOF cameras. We then propose a noise model which is able to explain some of the phenomena observed in the experiments. The model is based on assuming a noise source that is correlated with the light source (shot noise) and an additional additive noise source (dark current noise). The model predicts well the dependency of the distance errors on the image intensity and the true distance at an individual pixel.**

## I. INTRODUCTION

The TOF cameras work with an active illumination (an array of LEDs in infrared) [1]. The emitted light is then reflected by the objects in the scene and sensed by a pixel array in the camera. Thereby the light is attenuated such that the signal from far objects is attenuated more than the signal reflected from a near object, and therefore a correction of this distance-dependent attenuation must be performed (see [2]). Since the camera also receives the ambient light from the scene, a narrow-band infrared filter is used so that the received signal is less affected by the perturbing ambient light.

The light source of the TOF camera we used [3] is an amplitude-modulated 20 MHz infrared light, which lasts for a time duration between 0.5 and 50 ms. Each pixel of the camera sensor receives this incoming light and produces an electric signal proportional to the instantaneous value of the 20 MHz envelope. This signal is sampled synchronously with the envelope four times per period and the four samples $A_1, A_2, A_3, A_4$ are the basis for the calculation of the amplitude *a* (from which the intensity image is obtained) and the phase shift φ (from which the distance image is computed) corresponding to that pixel (for details see ref. [4],[5]).

In practice, however, these signals are affected by systematic errors. In fact one is faced with the quantum character of the infrared active illumination especially at low intensities.

## II. EXPERIMENTS

In the following we describe three experiments, which each puts in evidence a particular feature of this new type of camera.

### A. Intensity-dependent variance of the noise

We first considered a static scene and recorded one hundred intensity images with a fixed set of camera parameters (exposure time 20ms) at a fixed distance to the objects (1 m).

Let $I_k(i,j)$ be the value of the pixel (i,j) in the *k*-th image and $I_k$ the whole *k*-th image. For the set of 100 images we computed the average pixel values

$$\overline{I(i,j)} = \frac{1}{N}\sum_k I_k(i,j) \qquad (1)$$

and the pixel variances

$$\sigma^2(i,j) = \frac{1}{N}\sum_k \left(I_k(i,j) - \overline{I(i,j)}\right)^2 \qquad (2)$$

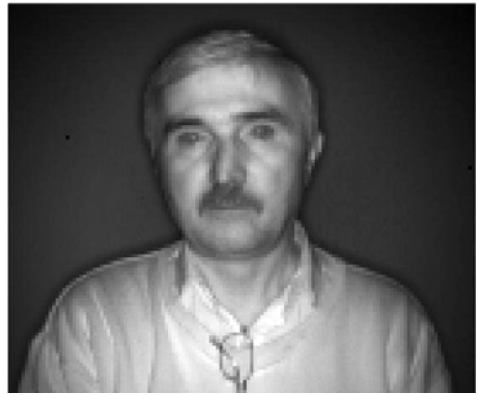

Figure 1a. The first picture from the 100 set.

In Fig. 1a we show an example image from the set of 100 images, in Fig. 1b the average image (note that the noise is reduced), and in Fig. 1c the square of standard deviation. It seems remarkable, though very natural for a Poisson process, that the mean and the standard deviation produce identical images (up to a scale factor). This result puts in evidence the fact that, indeed, our signals are generated in agreement to the well known physic laws (Poisson distribution of the shot noise). Nevertheless, images like those in Fig. 1c seem to have been obtained for the first time in the context of image processing.

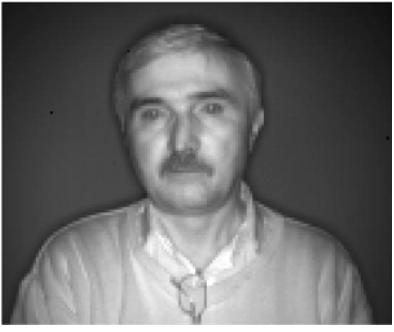

Figure 1b. The picture of the mean of a 100 pictures set.

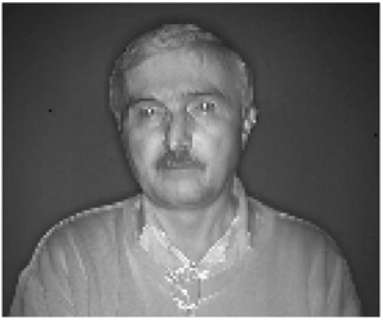

Figure 1c. The picture of the standard deviation of a 100 pictures set.

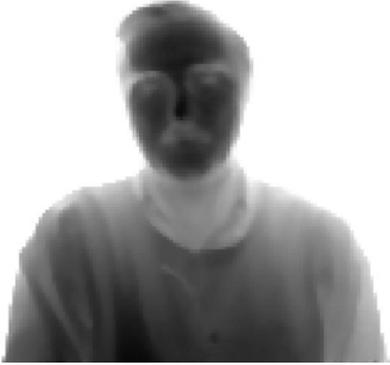

Figure 2. Distance image of a face. We notice that the moustache and the eyebrows are whiter, hence farther than the rest of the face.

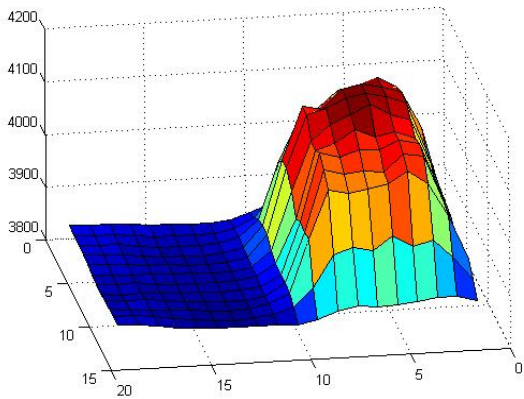

Figure 3. The 3D figure of a sheet of paper, half white and half black.

## B. Intensity-dependent distance error

A second set of experiments were addressing the distance measurements based on φ. One has the simple relationship (see [3]).

$$d = \frac{c \cdot T \cdot \varphi}{4 \cdot \pi} \quad (3)$$

between the distance d, and the phase shift φ (here c is the speed of light and T is the integration time).

We first noted that in the distance image of scene in the Fig. 2, the black moustache and the eyebrows was displaced about 2 cm behind the true distance. We then used a sheet of paper half white, half black at a distance of 4 m in front of the camera. Fig. 3 shows the corresponding distance image. Note the difference of more than 20 cm between the two halves. However, this error is not constant but depends on the distance and we shall describe the more complex dependencies next. At this point we shall only mention a few more measurements we made to clarify some dependencies.

First of all, the normal approximation for the Poisson law was evidenced for the measurements we made. Secondly, the linear dependency of image intensity on integration time is verified for objects at various distances (see Fig. 4) whereas the dependency of the distance signal on integration time was measured also for objects at various distances. Thereby one could infer the dependency of the distance signal on the intensity signal, a result which shows that at small intensity, the measured distance is incorrect (Fig. 6). This error, however, depends on the distance at which the objects are placed and it can be positive or negative.

## C. Surround-dependent distance errors

As shown in Fig. 6, the intensity-dependent error at a given pixel will also depend on the intensities of the surrounding pixels. The line marked with asterisks in Fig. 6 represents the measured distance to a white paper with a black background situated at a distance of 4 m. The line marked with diamonds represents the distance measured to the same white paper but with a white background. We observe that the background intensity change the measured distance with about 0.4 m. The distance error in Fig. 2 almost vanishes if the background is black. This shows that the difference between the results is very important not only for low intensities. We have, however, not yet further investigated this effect.

## III. NOISE MODEL FOR THE INTENSITY-DEPENDENT ERRORS

The amplitude $a$ and the phase $\varphi$ of the detected light are computed from the four samples A1, A2, A3, and A4 as

$$\varphi = \operatorname{atan}\left(\frac{A_4 - A_2}{A_1 - A_3}\right); \quad (4)$$

$$a = \frac{\sqrt{(A_4 - A_2)^2 + (A_1 - A_3)^2}}{2} \quad (5)$$

Knowing the light speed *c* and the period *T* of the modulation, the distance to the object, which reflected the light, is computed with relation (3); for more details se ref. [1], [3], [4] and [5].

We now propose a statistical model that matches the measured probability density functions of *d* and *a*.

We use a Monte Carlo method and we presume that the measured samples are modeled as

$$A_1 = g_1 \cdot (e_1 + n_1)$$
$$A_2 = g_2 \cdot (e_2 + n_2)$$
$$A_3 = g_3 \cdot (e_3 + n_3)$$
$$A_4 = g_4 \cdot (e_4 + n_4)$$
(6)

The $g_1, g_2, g_3, g_4$ represent the gains that are assumed to be equal to $g$. $e_1, e_2, e_3, e_4$ are the numbers of electrons produced by the detection of the incoming photons, and $n_1, n_2, n_3, n_4$ represent an additive noise introduced by the sensor, independent of the photon shot noise. We assume that these $n_i$ are normal random variables with mean $\bar{n}$ and standard deviations equal to $\sigma(n)$. In this relation $n_i$ is measured in the same units as $e_i$. The result of the simulation is not dependent on $\bar{n}$ since the mean will cancel out in Eqs. (1) and (2).

The $e_1, e_2, e_3, e_4$ are Poisson random variables with the means $\bar{e_1}, \bar{e_2}, \bar{e_3}, \bar{e_4}$ given by

$$\bar{e_1} = \bar{e} \cdot \left(\frac{1}{2} + \frac{\cos\varphi}{\pi}\right) \cdot T \qquad \bar{e_3} = \bar{e} \cdot \left(\frac{1}{2} - \frac{\cos\varphi}{\pi}\right) \cdot T$$
$$\bar{e_2} = \bar{e} \cdot \left(\frac{1}{2} - \frac{\sin\varphi}{\pi}\right) \cdot T \qquad \bar{e_4} = \bar{e} \cdot \left(\frac{1}{2} + \frac{\sin\varphi}{\pi}\right) \cdot T$$
(7)

and standard deviations given by:

$$\sigma_{e1} = \sqrt{\bar{e_1}} \qquad \sigma_{e3} = \sqrt{\bar{e_3}}$$
$$\sigma_{e2} = \sqrt{\bar{e_2}} \qquad \sigma_{e4} = \sqrt{\bar{e_4}}$$
(8)

where $\bar{e}$ represents the mean number of electrons produced by the light per unit of time in one pixel, and $T$ is the integration time.

To verify our model, we compute $\bar{d}(T)$, $\bar{a}(T)$, $\sigma_a(T)$, and $\sigma_d(T)$ as a function of the integration time, and compare these simulation results with the measured data.

To do so, we must know the values of $g$ and $\bar{e}$. These values can be deducted from the experimental data, assuming an integration time *T* that fulfills the condition $\bar{e} \cdot T \gg \bar{n}$. We use Eq. (2) and the relations for Poisson processes:

$$\sigma_a(T) = g \cdot \frac{\sqrt{\bar{e} \cdot T}}{\pi}$$
(9)

$$\bar{a}(T) = \frac{\bar{e} \cdot T \cdot g}{\pi}$$
(10)

to determine

$$\bar{e} = \left(\frac{\bar{a}(T)}{\sigma_a(T)}\right)^2 \frac{1}{T}$$
(11)

and

$$g = \pi \cdot \frac{(\sigma_a(T))^2}{\bar{a}(T)}$$
(12)

For the noise *n* we used a value of σ(*n*) equal to 43 electrons obtained from fitting the experimental to the simulated curve.

IV. RESULTS

The empirical results below are estimated for the central 3 by 3 pixels and 100 images of the same object, i.e. all the means are means of these 900 pixel values, and all the variances are variances of these 900 pixel values.

In Fig. 4 we present the results for the dependency of image intensity on integration time. The simulated results (continuous line) show the intensity *a* as a function of *Ti*. The experimental data measured for a white paper on a dark background are plotted with diamonds, and those measured for a white paper on a white background with asterisks. We observe a linear dependency and a very good match between the simulated and the measured data.

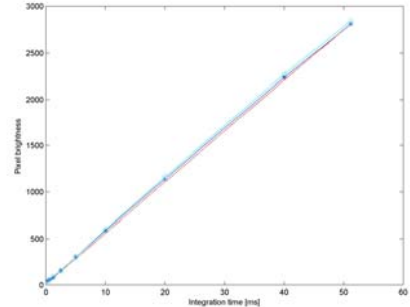

Figure 4. The intensity function of the integration time.

In Fig. 5 we present the results for the dependency of image-intensity variance as a function of integration time. We use the same plotting conventions as in the previous figure. Note that the model fits well the data measured on the white paper with black background.

In Fig. 6 we present the dependency of the distance on the image intensity at a fixed distance of four meters, and with the same objects and plotting conventions as before. First note that the errors are large for small intensities and that the model makes a good qualitative prediction of these errors. However, the surround intensities affect the distance measurement and the sign of the error, an effect that is currently not modeled.

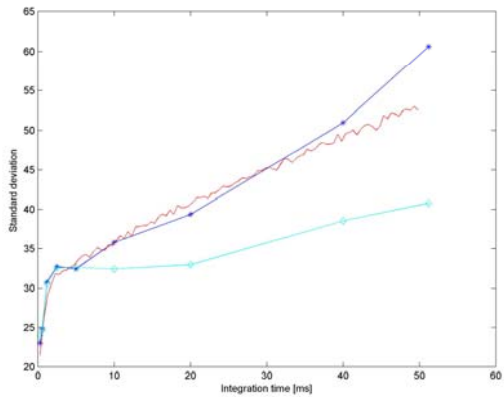

Figure 5. The standard deviation of the brightness function of the integration time.

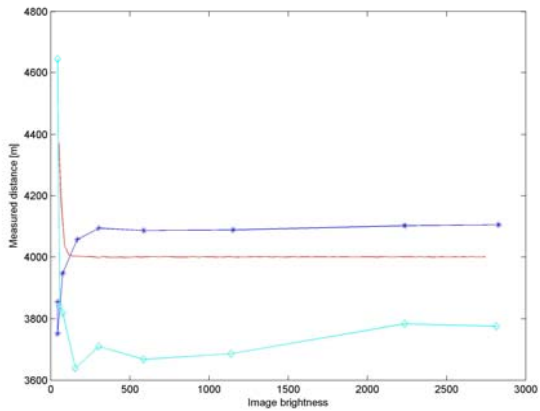

Figure 6. The measured distance function of the image brightness.

Our model shows that the systematic distance error measurements for dark objects and low image intensity is caused by *n*, the "dark current noise" of the camera. With *n=0*, the distance error vanishes in our simulation.

In Fig. 7 we simulate how the difference between a distance measured with an integration time of 0.1 ms and one measured with 50ms depends on the distance of the object relative to the camera (in the previous figure this distance was constant at four meters). Note that the error is positive in the range from 0m to 1.875m. At the distance of 1.875m the error is canceled. In the region from 1.875m to 3.75m the measured distance to a black object is smaller than to a white one, we can see that this is a periodic function.

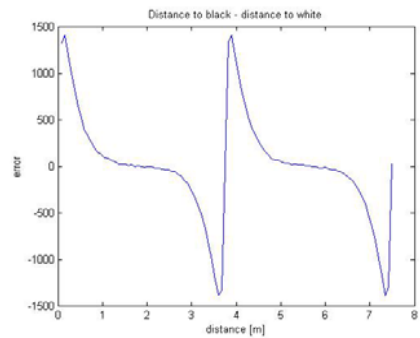

Figure 7. Distance-estimation errors as a function of the true distance

## V. CONCLUSIONS

Our main contribution is a model that predicts how the distance error at one pixel depends on image intensity at that pixel and on the distance itself. Qualitatively, darker objects will seem to be either further away or closer depending on the distance; an effect that is due to the nonlinearities in the phase-shift estimation. Our model, however, does not explain the influence of the surround intensity on the local pixel values, which remains a topic of future investigation since these effects are significant. The error caused by the surround intensity maybe due to hallow, smearing, multiple reflections of the light in the camera body or by the illumination of the pixel image by indirect light.

### ACKNOWLEDGMENT

The ARTTS project is funded by the European Commission (contract no. IST-34107) within the Information Society Technologies (IST) priority of the 6th Framework Programme.

This publication reflects the views only of the authors, and the Commission cannot be held responsible for any use which may be made of the information contained therein.